\documentclass[doublecol]{epl2} 
\usepackage{amssymb, amsmath, amsthm}

\usepackage{graphicx}
\usepackage{dcolumn}
\usepackage{bm}

\usepackage{dsfont}
\usepackage{color}
\usepackage{hyperref}%
\usepackage{esint}
\usepackage{float}

\title{Full-counting statistics for the alternating and domain-wall states}
\shorttitle{Full-counting statistics for the alternating and domain-wall states} 

\author{Oleksandr Gamayun}
\shortauthor{O. Gamayun}

\institute{                    
London Institute for Mathematical Sciences, Royal Institution,
	21 Albemarle St, London W1S 4BS, UK
}

\abstract{
For a one-dimensional system of free fermions, we derive a connection between the full counting statistics of domain-wall and alternating occupancy (Néel)  states. This allows us to demonstrate asymptotic linear growth with time of the even cumulants in the Néel state.}

\begin{document}

\maketitle

The quantum evolution of highly non-equilibrium states attracts a lot of attention \cite{RevModPhys.83.863}. 
For one-dimensional systems, the effects of the constrained dynamics are especially pronounced, both in transport and statistical properties \cite{Ilievski2016,Vasseur2016}.
Inspired by ultra-cold atom experiments \cite{wienand2023emergence} a recent work \cite{fujimoto2024exact} exactly calculated the variance
of a bipartite fluctuation in one-dimensional noninteracting fermionic dynamics starting from an
alternating (A) occupancy state . 
This quantity grows linearly at long times, which starkly contrasts with the dynamics for the same system initialized in the domain-wall (DW) state \cite{Moriya2019,10.21468/SciPostPhys.8.3.036}.  
Surprisingly, the averaged number of particles \cite{PhysRevE.78.061115,PhysRevE.59.4912,Viti2016,Moriya2019,10.21468/SciPostPhys.8.3.036} in the DW case coincides with the variance in the A case given in \cite{fujimoto2024exact}
(after some rescalings,  see Eq. \eqref{C1}). 

In this note, we extend this relation to all even cumulants of the bipartite fluctuations. 
More precisely, we consider a lattice system of free fermions with $4L$ sites, whose dynamics is described by the hopping Hamiltonian 
\begin{equation}\label{H}
	H = - \frac{1}{2} \sum\limits_{i=1}^{4L} \left(a_{i+1}^+a_i + a_{i}^+a_{i+1}\right). 
\end{equation}
We assume hard-wall boundary conditions, $a_{4L+1}=0$, $a_{-1} =0$. 
We specify by $N_R = \sum\limits_{i=2L+1}^{4L}a_i^+a_i$ the number of particles in the right part of the system. 
We are going to study the cumulant-generating function, also known as the full-counting statistics (FCS)
\begin{equation}
	\chi_\Omega(\lambda,t) = \langle\Omega | e^{\lambda N_R(t)}
	|\Omega \rangle \equiv \langle\Omega | e^{itH}e^{\lambda N_R}
	e^{-itH} |\Omega \rangle. 
\end{equation}
The cumulants  can be extracted from the FCS as coefficients for the expansion at $\lambda=0$
\begin{equation}\label{chi0}
	\ln \chi_\Omega(\lambda) = \lambda C^{(1)}_\Omega+ \sum\limits_{k=2}^\infty \frac{\lambda^{k}}{k!} C_\Omega^{(k)},
\end{equation}
with $C^{(1)}_\Omega = \langle\Omega |N_R(t)|\Omega \rangle$.

The initial state $|\Omega \rangle$ for the DW case is given by $|{\rm DW}\rangle= \prod\limits_{i=1}^{2L}|i\rangle$, while for $A$-state (Néel) we specify the state as 
$|{\rm A}\rangle= \prod\limits_{i=1}^{2L}|2i\rangle$. 
Here $|i\rangle = a_i^+ |0\rangle$ is a single fermion state in the site $i$ \footnote{The vacuum is a state such that $a_k|0\rangle=0$, $\forall k$}. 
One can also use a slightly different definition of these states $|{\rm \widetilde{ DW}}\rangle= \prod\limits_{i=2L+1}^{4L}|i\rangle$ and $|\widetilde{{\rm A}}\rangle= \prod\limits_{i=1}^{2L}|2i-1\rangle$.
Because of the particle-hole symmetry the corresponding statistics are closely related
\begin{equation}\label{rel}
	\frac{\chi_{\rm A}(-\lambda,t)}{ \chi_{\widetilde{\rm A}}(\lambda,t) }  =  \frac{\chi_{\rm DW}(-\lambda,t)}{ \chi_{\widetilde{\rm DW}}(\lambda,t) }  = e^{-2\lambda L}.
\end{equation}

We now formulate the main statement of this note
\begin{equation}\label{main}
	\boxed{\chi_{\rm A}(\lambda,t)\chi_{\rm A}(-\lambda,t) = \chi_{\rm DW}(2\ln\cosh(\lambda/2),2t).}
\end{equation}

This relation, together with \eqref{chi0}, immediately allows us to connect different cumulants for the two initial states. The first few such connections are 
\begin{align}
	\label{C1} 4C_{\rm A }^{(2)}(t) &= C^{(1)}_{\rm DW}(2t),\\
	4C_{\rm A }^{(4)}(t) &= 3C^{(2)}_{\rm DW}(2t)-C^{(1)}_{\rm DW}(2t),\\
	16C_{\rm A }^{(6)}(t) &= 4C^{(1)}_{\rm DW}(2t)-15 C^{(2)}_{\rm DW}(2t).
\end{align}
The full expression for the FCS in the DW case is available in the form of a Fredholm determinant of Bessel kernel type in Refs. \cite{Moriya2019,10.21468/SciPostPhys.8.3.036}. In this form, its long-time asymptotic can be obtained analytically \cite{Bothner2019}
\begin{equation}
	\ln\frac{\chi_{\rm DW}(\lambda,t)}{G^2\left(1+\frac{i\lambda}{2\pi}\right)G^2\left(1-\frac{i\lambda}{2\pi}\right)} = \frac{\lambda t}{\pi} + \frac{\lambda^2\ln(4t)}{2\pi} + o(1).
\end{equation}
Here $G(x)$ is the Barnes G-function. 
Combining this expression with \eqref{main} and \eqref{chi0} one immediately gets the linear growth of all even cumulants in the A-case. Moreover, even cumulants completely define the entanglement entropy $S_{\rm A}$ between the two parts of the system \cite{PhysRevLett.102.100502,PhysRevB.83.161408,Song2012}. This way, we obtain the linear growth of entanglement
\begin{equation}
	S_{\rm A}(t) = \int\limits_{-\infty}^\infty \frac{\ln \chi_{\rm A}(\lambda,t)}{4\sinh^2(\lambda/2)} d\lambda \overset{t\to \infty}{=} \frac{2\ln 2}{\pi}t.
\end{equation}

To prove \eqref{main} we first introduce notations for the eigenvectors of the Hamiltonian \eqref{H} 
$H|\alpha\rangle= \varepsilon_\alpha |\alpha\rangle$. We parametrize them by the same integers as the lattice sites $\alpha = 1 ,2, \dots 4L$ but use Greek letters to avoid confusion.
The corresponding energies are $\varepsilon_\alpha = \cos\left(\pi \alpha/(4L+1)\right)$
and explicit expressions for $|\alpha\rangle$ reads 
\begin{equation}\label{eigen}
	|\alpha\rangle  = \sqrt{\frac{2}{4L+1}}\sum\limits_{j=1}^{4L} \sin \left(\frac{\pi \alpha j}{4L+1}\right)|j\rangle.
\end{equation}
The FCS can be computed employing Wick’s theorem for group-like elements of the fermionic algebra \cite{Alexandrov2013}. In the DW case this leads to 
\begin{equation}
	\chi_{\rm DW}(\lambda,t) = \det_{1\le j, k \le 2L} \left(\delta_{jk} +z_\lambda X_{jk}\right)
\end{equation}
where $z_\lambda = (e^{\lambda}-1)$ and the matrix $X$ after the insertions of the resolutions of unity in the energy space reads
\begin{equation}\label{X}
	X_{jk} = \sum_{\alpha,\beta} \langle j|\alpha\rangle e^{it\varepsilon_\alpha}\langle \alpha| P_R| \beta\rangle e^{-it\varepsilon_\beta}\langle \beta| k\rangle.
\end{equation}
Here $P_R$ is a projector on the right-hand side of the system. It can be written as $P_R = |{\rm \widetilde{ DW}}\rangle \langle \widetilde{ DW}|$ understood in the single particle sector. The corresponding matrix in the energy space is denoted as $\mathcal{R}$ with matrix elements 
\begin{equation}\label{single}
	\mathcal{R}_{\alpha\beta}\equiv\langle \alpha| P_R| \beta\rangle = \sum_{k=2L+1}^{4L} \langle \alpha| k\rangle \langle k| \beta\rangle.
\end{equation}
We can rewrite the FCS as a determinant in the energy space. Understanding sums in \eqref{X} as matrix multiplication and using the cyclic relation $\det (1 + AB) = \det(1+BA)$ we arrive at 
\begin{equation}\label{chi}
	\chi_{\rm DW}(\lambda,t) = \det_{1\le n,m\le 4L} \left(1 +z_\lambda\mathcal{R}_t \mathcal{L} \right). 
\end{equation}
Here $\mathcal{R}_t = \Lambda_t \mathcal{R} \Lambda^{-1}_t $  stands for the time dependent projector; 
by $\Lambda_t$ we have denoted the diagonal matrix with the entries $(\Lambda_t)_{\alpha\beta}= e^{it\varepsilon_n}\delta_{\alpha\beta}$. Additionally, we have introduced 
$\mathcal{L}_{\alpha\beta} = \langle \alpha|{\rm  DW}\rangle \langle {\rm DW}|\beta\rangle$, understood in the same sense as \eqref{single}. Obviously $\mathcal{R}+\mathcal{L}=1$.

In Eq. \eqref{chi} one immediately sees that the factor $\mathcal{R}_t$ is responsible for the observable and $\mathcal{L}$ for the initial state. 
For the A-state instead of the matrix $\mathcal{L}$ we have a much simpler expressions:
$\langle \alpha|{\rm  A}\rangle \langle {\rm A}|\beta\rangle = (\delta_{\alpha\beta} -\mathcal{P}_{\alpha\beta})/2$ 
with $\mathcal{P}_{\alpha\beta} = \delta_{\alpha,4L+1-\beta}$. This can be obtained using the exact form of the eigenvectors \eqref{eigen} or making use of the chiral symmetry present in the system. In the single-particle basis the symmetry is given by $C= \mathds{1}_{2L}\otimes \sigma_z$, satisfying $\{ H,C \}=0$. One can easily see that $C$ distinguishes odd and even sites. Moreover, the way we have parametrized energy allows us to conclude $\varepsilon_{4L+1-\alpha} = - \varepsilon_{\alpha}$, which is equivalent to $\mathcal{P} \Lambda_t= \Lambda^{-1}_t\mathcal{P}$. This relation together with $\mathcal{P}^2 =1 $ and $\mathcal{P}\mathcal{R} = \mathcal{R}\mathcal{P} $ leads to 
\begin{equation}\label{RR}
	(\mathcal{R}_t\mathcal{P})^{2} = \Lambda^{-1}_{t}\mathcal{R}_{2t}\mathcal{R}\Lambda_t.
\end{equation}
Taking into account that $\mathcal{R}_t$ is  a projector i.e. $\mathcal{R}_t^2 = \mathcal{R}_t$ we can present 
\begin{multline}\label{DW}
	\chi_{\rm DW}(\lambda,t) = \det \left(1 +z_\lambda\mathcal{R}_t (1-\mathcal{R}) \right) = \\
	\det \left(1 +z_\lambda\mathcal{R}_t \right)\det \left(1 + z_{-\lambda}\mathcal{R}_t\mathcal{R} \right)
\end{multline}
Notice that the first determinant can be easily computed 
$\det \left(1 +z_\lambda\mathcal{R}_t \right) =(1 +z_\lambda)^{2L}  = e^{2 \lambda L }$. 
Similary for the alternating state $|{\rm A}\rangle$ we obtain
\begin{multline}\label{detA}
	\chi_{\rm A}(\lambda,t) = \det \left(1 +z_\lambda\mathcal{R}_t \frac{1-\mathcal{P}}{2} \right) = \\
	\det \left(1 +z_\lambda\mathcal{R}_t/2 \right)\det \left(1 - \mu_{\lambda}\mathcal{R}_t\mathcal{P} \right)
\end{multline}
where $\mu_\lambda = (e^\lambda -1)/(e^\lambda +1)$.
We notice that 
\begin{multline}
	\det \left(1 - \mu_{\lambda}\mathcal{R}_t\mathcal{P} \right)\det \left(1 + \mu_{\lambda}\mathcal{R}_t\mathcal{P} \right) \\ = \det \left(1 - \mu^2_{\lambda}(\mathcal{R}_t\mathcal{P})^2 \right) = \det(1-\mu^2_{\lambda} \mathcal{R}_{2t}\mathcal{R} ),
\end{multline}
where in the last step we have used \eqref{RR} and eliminated the matrices $\Lambda_t$ 
under the determinant. At the final step we use 
\begin{equation}
	\det \left(1 +z_\lambda\mathcal{R}_t/2 \right) =  e^{\lambda L}\left(\cosh(\lambda/2)\right)^{2L},
\end{equation}
to arrive that the result  \eqref{DW}. Moreover, from the relations \eqref{DW} and \eqref{detA}, Eq.  \eqref{rel} follows immediately.



\acknowledgments
I am grateful to F. C. Sheldon and A. Stepanenko for illuminating discussions,  careful reading of the manuscript and numerous useful remarks and suggestions.  I thank K. Fujimoto for pointing out mistakes in the first version. 

\bibliographystyle{eplbib}
\bibliography{litra}

\end{document}